\documentclass[conference]{IEEEtran}
\usepackage{caption}
\usepackage{cite}
\usepackage{amsmath,amssymb,amsthm,mathtools}
\usepackage{bbm}
\usepackage{graphicx}
\usepackage{enumitem}
\usepackage{booktabs}
\usepackage{tikz}
\usetikzlibrary{trees,fit,positioning,arrows.meta}

\newcommand{\bc}{\mathbf{c}}
\newcommand{\bu}{\mathbf{u}}
\newcommand{\bx}{\mathbf{x}}
\newcommand{\by}{\mathbf{y}}
\newcommand{\bz}{\mathbf{z}}
\newcommand{\F}{\mathbb{F}}

\newcommand{\Iset}{\mathcal{I}}

\newcommand{\C}{\mathcal{C}}

\newcommand{\Mset}{\mathcal{M}}
\newcommand{\Ir}{\mathcal{I}_r}
\newcommand{\wt}{\mathrm{w}}

\newcommand{\Ev}{\operatorname{ev}}

\newcommand{\ind}{\operatorname{ind}}

\newcommand{\wmin}{w_{\min}}
\newcommand{\UB}{\mathrm{UB}}

\newtheorem{definition}{Definition}
\newtheorem{lemma}{Lemma}

\newtheorem{theorem}{Theorem}

\begin{document}

\title{A Hybrid Reliability-Weight Framework for Construction of Polar Codes}

\author{
\IEEEauthorblockN{Mohammad Rowshan}
\IEEEauthorblockA{University of New South Wales\\
Sydney, Australia\\
Email: m.rowshan@unsw.edu.au}
\and
\IEEEauthorblockN{Vlad-Florin Dr\u{a}goi}
\IEEEauthorblockA{Aurel Vlaicu University\\
Arad, Romania\\
Email: vlad.dragoi@uav.ro}
}

\maketitle

\begin{abstract}
Polar codes are usually constructed by ranking synthetic bit-channels according to reliability, which guarantees capacity-achieving behavior but can yield poor low-weight spectra at short and moderate lengths. Recent algebraic results express the contribution of individual bit-channels to the multiplicities of minimum and near-minimum weight codewords in closed form. In this work we combine these insights into a mixed (reliability--weight) bit-channel ordering. We define a per-bit cost whose distance term is derived from orbit enumeration of minimum-weight codewords and scaled by a Bhattacharyya-type factor, and show that the resulting mixed construction minimizes a truncated SC/ML union-bound surrogate within a class of decreasing monomial codes. We relate the mixed metric to error events in SCL decoding via a pruning/ML decomposition, and prove that mixed designs act as local perturbations of reliability-based constructions whose asymptotic impact vanishes as code-length approaches infinity. 
Numerical results for short and moderate lengths on BPSK-AWGN, implemented via Gaussian approximation and closed-form weight contributions, illustrate the trade-off between pure reliability-based and mixed constructions in terms of minimum distance, multiplicity, and union-bound approximations. All proofs are deferred to the appendices.
\end{abstract}

\section{Introduction}

Polar codes~\cite{Arikan} achieve the capacity of any binary-input memoryless symmetric (BMS) channel with low-complexity encoding and successive-cancellation (SC) decoding. Practical constructions order synthetic channels by reliability, obtained via density evolution (DE), Gaussian approximation (GA)~\cite{mori,chung,trifonov}, or Tal-Vardy construction \cite{tal2}. At short and moderate blocklengths, however, such purely reliability-based designs may have poor low-weight spectra, making them suboptimal under maximum-likelihood (ML) or list decoding compared to, e.g., Reed--Muller codes. Many search methods for good polar codes exist, but they lie outside the scope of this work due to their  high computational cost.

The algebraic description of polar codes as \emph{decreasing monomial codes}~\cite{bardet} enables closed-form expressions for the minimum distance and the multiplicity of minimum (and some near-minimum) weight codewords via the action of the lower triangular affine group~\cite{dragoi17thesis,vlad1.5d,ye2024distribution,rowshan2025weight,dragoi2025polar}. These formulas naturally assign to each maximum-degree monomial a weight-contribution score measuring its impact on the minimum-weight spectrum, and have been used to propose weight-aware partial orders for low-rate polar-like codes~\cite{rowshan2025towards}.

Most practical constructions use a \emph{universal} reliability sequence~\cite{schurch,mondelli-construction,he}. This sequence can be improved considering the weight contribution of bit-channels as in~\cite{rowshan2025towards}.  This work raises the question of how to systematically combine reliability and distance information in a bit-wise design metric, especially for near-ML decoders such as SCL \cite{tal2015list}. A code with good distance properties can then be further improved by outer concatenation with CRC \cite{niu2012crc}, convolutional pretransform in PAC coding \cite{arikan2,rowshan-pac1} and profile shifted PAC  (PS-PAC) coding \cite{gu2025pac}. 

In this work we use the decreasing-monomial representation and closed-form expressions for the minimum distance $\wmin$ and its multiplicity $A_{\wmin}$ (via orbits of maximum-degree monomials) to define a $K$-dependent mixed per-bit cost that blends SC-based reliability with an orbit-based distance penalty. We interpret this score as minimising a scalarised SC/ML functional derived from a truncated union bound, show that sub-maximal degrees are asymptotically negligible so only maximum-degree monomials require a distance term, and relate the metric to SCL via a pruning/ML error decomposition. Mixed designs then appear as local perturbations of reliability-based constructions whose impact on SC and ML union bounds vanishes as $N\to\infty$, while numerical examples on BPSK-AWGN show clear finite-length gains in $\wmin$ or $A_{\wmin}$ for only modest reliability loss.


\section{Preliminaries and Structural Tools}
\label{sec:prelim}

\subsection{Notation and union bound}

Let $\F_2$ be the binary field and $N=2^m$ the blocklength. Vectors in $\F_2^N$
are in bold, with Hamming weight $\wt(\bc)$. A binary linear code
$\C\subseteq\F_2^N$ has minimum distance
\[
  \wmin(\C) = \min\{\wt(\bc): \bc\in\C\setminus\{0\}\},
\]
and weight enumerator $A_w = |\{\bc\in\C:\wt(\bc)=w\}|$. For a BMS channel $W$
and ML decoding, the union bound is
\begin{equation}
  P_B^{\mathrm{ML}}(\C,W)
  \le
  \sum_{w=\wmin}^N A_w\,P_w(W),
  \label{eq:ML-union-bound}
\end{equation}
where $P_w(W)$ is the pairwise error probability against any weight-$w$
competitor. For many BMS channels there exist constants $c(W)>0$ and
$\gamma(W)\in(0,1)$ such that $P_w(W)\le c(W)\gamma(W)^w$ for all $w$.

\subsection{Polar codes and bit-channel reliabilities}

Polar codes of length $N=2^m$ are generated by $G_N = G_2^{\otimes m}$ with
$G_2=\bigl[\begin{smallmatrix}1&0\\1&1\end{smallmatrix}\bigr]$~\cite{Arikan}.
For a BMS channel $W$, polarization induces synthetic channels
$W_N^{(i)}$ with Bhattacharyya parameters $Z_i=Z(W_N^{(i)})$ or estimated SC
bit-error probabilities $P_i^{\mathrm{SC}}$~\cite{MoriTanaka}. A classical
construction chooses an information set $\Iset$ of size $K$ by taking the $K$
indices with smallest $Z_i$. Under SC decoding,
\begin{equation}
  P_B^{\mathrm{SC}}(\C(\Iset),W)
  \le
  \sum_{i\in\Iset} Z_i,
  \label{eq:SC-bler-bound}
\end{equation}
which motivates minimising $\sum_{i\in\Iset} Z_i$.

\subsection{Monomial representation and decreasing codes}

Let $\bx=(x_0,\dots,x_{m-1})$ and
$\mathcal{R}_m = \F_2[x_0,\dots,x_{m-1}]/(x_i^2-x_i)_{i=0}^{m-1}$. A monomial
is $f=x_{i_1}\cdots x_{i_s}$ with $0\le i_1<\dots<i_s$; its degree is
$\deg(f)=s$ and its index set is
$\ind(f)=\{i_1,\dots,i_s\}\subseteq\{0,\dots,m-1\}$. Let $\Mset_m$ be the set
of all such monomials. Evaluation on all $\bz\in\F_2^m$ gives
$\Ev(f)=(f(\bz))_{\bz\in\F_2^m}\in\F_2^N$.

There is a standard bijection between indices $i\in\{0,\dots,N-1\}$ and
monomials: for $f=\prod_{t\in S}x_t$, define $b_t=1$ if $t\in S$ and $b_t=0$
otherwise, and set $i(f)=\sum_{t=0}^{m-1} b_t 2^t$. Polar and Reed--Muller
codes are monomial codes under this mapping~\cite{SasogluMonomial}.

\begin{definition}[Monomial and decreasing monomial codes]
For $\Iset\subseteq\Mset_m$, the associated monomial code is
$\C(\Iset)=\operatorname{span}\{\Ev(f):f\in\Iset\}\subseteq\F_2^N$.
A partial order $\preceq$ on $\Mset_m$ is defined by monomial divisibility and
degree-preserving ``shifts''~\cite{SasogluMonomial}. A set
$\Iset\subseteq\Mset_m$ is \emph{decreasing} if $f\in\Iset$ and $g\preceq f$
imply $g\in\Iset$. The code $\C(\Iset)$ with $\Iset$ decreasing is called a
\emph{decreasing monomial code}.
\end{definition}

Under the index--monomial mapping, both polar codes and Reed--Muller codes
correspond to decreasing sets.

\subsection{Minimum distance and orbit-based $A_{\wmin}$}

Let $r=\max_{f\in\Iset}\deg(f)$ be the maximum degree in $\Iset$, and
$\Ir(\Iset)$ the subset of degree-$r$ monomials. Then the minimum distance of
$\C(\Iset)$ depends only on $r$ and is given by~\cite{SasogluMonomial,DragoiPhD}
\begin{equation}
  \wmin(\Iset) = 2^{m-r}.
  \label{eq:wmin-formula}
\end{equation}
All minimum-weight codewords arise from the orbits of degree-$r$ monomials under the lower triangular affine group $\mathrm{LTA}(m,2)$. 
For $f=x_{i_1}\cdots x_{i_r}\in\Ir$ with $i_1<\dots<i_r$, define
\(
  |\lambda_f(x_{i_\ell})|
  \triangleq
  |\{j\in[0,i_\ell-1]: x_j\notin\ind(f)\}|,
\)
\[
  |\lambda_f| \triangleq \sum_{\mathclap{x_i\in\ind(f)}}|\lambda_f(x_i)|.
\]
Then the orbit of $f$ has size $2^{r+|\lambda_f|}$, and summing over
$f\in\Ir(\Iset)$ yields the minimum-weight multiplicity
\begin{equation}
  A_{\wmin}(\Iset)
  = \sum_{f\in\Ir(\Iset)} 2^{r+|\lambda_f|},
  \label{eq:Awmin-ir-sum}
\end{equation}
which depends only on the subset $\Ir(\Iset)$.

This naturally suggests assigning to each degree-$r$ monomial a
\emph{minimum-weight contribution} $\mathrm{contrib}_{\wmin}(f) = 2^{r+|\lambda_f|}$.

\section{Mixed Reliability--Weight Design}
\label{sec:mixed}

We now introduce a $K$-dependent mixed metric combining bit-channel reliability
with a distance penalty derived from~\eqref{eq:Awmin-ir-sum}. Throughout this
section we fix a length $N=2^m$, a dimension $K$, a BMS channel $W$, and a
decreasing monomial code $\C(\Iset)$ of size $K$.

\subsection{Bit-wise ML contribution and local distance}

We first recall a coset-type notion of distance at the bit level.

\begin{definition}[Bit-wise minimum distance and ML contribution]
Let $\C\subseteq\{0,1\}^N$ be a binary code and $i\in\{0,\dots,N-1\}$. The
\emph{bit-wise minimum distance} of position $i$ is
\[
  w_{\min}(i)
  =
  \min\{\wt(\bc): \bc\in \C,\ c_i=1\},
\]
with the convention $w_{\min}(i)=+\infty$ if no such codeword exists.

For a fixed BMS channel $W$, let $P_w(W)$ denote the pairwise error
probability of any weight-$w$ codeword. The \emph{bit-wise ML contribution} of
index $i$ is
\[
  \Phi^{\mathrm{ML}}(i)
  =
  \sum_{w\ge w_{\min}(i)} A_w^{(i)}\,P_w(W),
\]
where $A_w^{(i)}$ are per-index multiplicities from a fixed partition of the
ML union bound~\eqref{eq:ML-union-bound}.
\end{definition}

Thus $w_{\min}(i)$ measures the minimum weight of any codeword involving index
$i$, and $\Phi^{\mathrm{ML}}(i)$ is the portion of the ML union bound that we
attribute to $i$.

\subsection{Monomial-level weight penalty at fixed $K$}

Fix a decreasing monomial set $\Iset$ of size $K$, with maximum degree
$r=\max_{f\in\Iset}\deg(f)$. In the class of decreasing sets with the same
maximum degree $r$, the minimum distance $\wmin=2^{m-r}$ is fixed, and only
$A_{\wmin}$ varies via the choice of $\Ir(\Iset)$.

\begin{definition}[Structural weight penalty]
For each index $i$ with associated monomial $f_i$, define
\begin{equation}
  C_K(i) =
  \begin{cases}
    2^{r+|\lambda_{f_i}|}, & \text{if }\deg(f_i)=r,\\[0.2em]
    0, & \text{if }\deg(f_i) \neq r,
  \end{cases}
  \label{eq:CK-def}
\end{equation}
so that $A_{\wmin}(\Iset)=\sum_{i\in\Iset} C_K(i)$.
\end{definition}

Thus, $C_K(i)$ is an per-bit weight contribution to the minimum-weight multiplicity. It is strictly increasing in $|\lambda_{f_i}|$; hence, consistent with the weight-contribution order in~\cite{RowshanDragoiWeightStructure}.

\subsection{Channel-dependent distance penalty and mixed cost}

For a fixed BMS channel $W$ we approximate the pairwise error probabilities in
the ML union bound by a Bhattacharyya-type exponential. In particular, we use
\[
  P_w(W) \approx Z(W)^w,
\]
where $Z(W)\in(0,1)$ is the Bhattacharyya parameter of the physical channel.
Under this approximation, the minimum-weight term in the union bound scales as
\[
  A_{\wmin} P_{\wmin}(W)
  \approx
  A_{\wmin} Z(W)^{\wmin}
  =
  \sum_{i\in\Iset} C_K(i)\,Z(W)^{\wmin}.
\]

This motivates weighting $C_K(i)$ by $Z(W)^{\wmin}$ and using the result as a
channel-dependent per-bit distance penalty.

\begin{definition}[Mixed monomial cost]
For each index $i$ define
\begin{equation}
  D_K(i)
  \triangleq
  C_K(i)\,Z(W)^{\wmin},
  \label{eq:DK-def}
\end{equation}
and, for a tuning parameter $\alpha>0$,
\begin{equation}
  J_K(i)
  \triangleq
  Z_i + \alpha\,D_K(i),
  \label{eq:JK-def}
\end{equation}
where $Z_i$ is any chosen reliability metric (e.g., GA-based $P_i^{\mathrm{SC}}$). The parameter $\alpha$ controls the \emph{relative} weight of SC-type and ML-type contributions so that the spread of $\alpha D_K(i)$ becomes comparable to or   larger than the spread of $Z_i$, so the mixed ordering meaningfully re-ranks degree-$r$ indices according to their minimum-weight contribution.
\end{definition}

For $\deg(f_i)<r$ we have $C_K(i)=0$, hence $D_K(i)=0$; lower-degree monomials
are ordered purely by reliability. For maximum-degree monomials, the
reliability and distance terms interact through $J_K(i)$.

\subsection{Mixed construction}

For decreasing monomial sets $\Iset$ with fixed maximum degree $r$, define the
mixed functional
\begin{equation}
  F_K(\Iset)
  \triangleq
  \sum_{i\in\Iset} J_K(i)
  =
  \sum_{i\in\Iset} Z_i
  + \alpha \sum_{i\in\Iset} D_K(i).
  \label{eq:FK-def}
\end{equation}
The first term approximates the SC union bound~\eqref{eq:SC-bler-bound},
while, under the Bhattacharyya approximation, the second term is proportional
to the minimum-weight contribution $A_{\wmin} Z(W)^{\wmin}$ in the ML union
bound.

\begin{definition}[K-dependent mixed design]
Let $\mathcal{D}_K^{(r)}$ be the family of decreasing monomial sets $\Iset$
with $|\Iset|=K$ and $\max_{i\in\Iset}\deg(f_i)\le r$. Any
\[
  \Iset_{\mathrm{mix}}(K)
  \in
  \arg\min_{\Iset\in\mathcal{D}_K^{(r)}} F_K(\Iset)
\]
is called a \emph{K-dependent mixed design} at maximum degree $r$.
\end{definition}

In the numerical implementation (Section~\ref{sec:numerical}), we will
instantiate $Z_i$ as GA-based SC error estimates and $Z(W)$ using a simple
closed-form approximation for BPSK-AWGN; $\alpha$ is treated as a tuning
parameter (e.g., a fixed constant such as $\alpha=100$) that controls the
relative emphasis on reliability vs distance.

\subsection{ML-negligibility of sub-maximal degrees}

We justify concentrating the distance penalty $D_K(i)$ on maximum-degree
monomials by showing that lower-degree positions become asymptotically
irrelevant in the ML union bound.

\begin{lemma}[Sub-maximal degrees are ML-negligible]
\label{lem:negligible-high-weight}
Let $W$ be a fixed BMS channel. Assume there exist constants $c(W)>0$ and
$\gamma(W)\in(0,1)$ such that
$P_w(W) \le c(W)\gamma(W)^w$ for all $w\ge1$. Let $N=2^m$ and
$\wmin=2^{m-r}$ be the minimum distance of a decreasing monomial code with
maximum degree $r$. If
\[
    \deg(f_i)\le r-1 \Rightarrow w_{\min}(i)\ge 2\wmin,
\]
\[
    \deg(f_j)=r \Rightarrow w_{\min}(j)=\wmin
\]
for some $j$, then
\[
  \frac{\Phi^{\mathrm{ML}}(i)}{\Phi^{\mathrm{ML}}(j)}
  \xrightarrow[m\to\infty]{} 0
  \quad\text{for all }i\text{ with }\deg(f_i)\le r-1.
\]
\end{lemma}

\noindent
Intuitively, indices of degree $\le r-1$ only participate in codewords of
weight at least $2\wmin$ and are penalised by an extra factor $\gamma(W)^{\wmin}$ compared to degree-$r$ indices. 

Lemma~\ref{lem:negligible-high-weight} supports focusing the distance penalty
on maximum-degree monomials $f_i$ and ranking sub-maximal degrees purely by
reliability.
\subsection{Design-SNR–Driven Polar–to–RM Distance Transition}
\begin{theorem}[Design-SNR and min-distance staircase] 
\label{thm:distance-staircase}
For $i\in\{0,\dots,N-1\}, N=2^m$, let $D_i$ be the Hamming weight of row $i$ of $G_N$ (equivalently $D_i=2^{\wt(i)}$). Assume that the design-SNR-dependent Bhattacharyya parameters of the synthetic channels have the form
\begin{equation}
  Z_i(\rho) = a_i e^{-c D_i \rho},
  \qquad a_i>0,\ c>0,\ \rho\ge0,
  \label{eq:exp-model}
\end{equation}
(e.g., exact for BEC and a standard exponential approximation for BPSK-AWGN).
For each $\rho\ge0$, let $\Iset_{\mathrm{rel}}(\rho,K)$ be the set of
$K$ indices with smallest $Z_i(\rho)$, 
and denote 
\[
  d(\rho)
  \triangleq
  \min_{i\in\Iset_{\mathrm{rel}}(\rho,K)} D_i,
  \qquad
  \wmin(\rho)
  \triangleq
  d(\rho).
\]

Then:
\begin{enumerate}[label=(\roman*),leftmargin=*]
\item For any $0\le\rho_1<\rho_2$, we have
      $\wmin(\rho_2)\ge\wmin(\rho_1)$. Moreover, $\wmin(\rho)$ is
      piecewise constant in $\rho$ with finitely  upward jumps; each jump
      occurs exactly when the last selected bit-channel of some row-weight
      $D_i=d$ leaves $\Iset_{\mathrm{rel}}(\rho,K)$, after which
      $\wmin(\rho)$ increases to the next larger row-weight present in
      $\{D_i : i\in\Iset_{\mathrm{rel}}(\rho,K)\}$.

\item Let $r^\star$ be the smallest integer such that
      $\sum_{t=0}^{r^\star}\binom{m}{t}\ge K$, and let
      $\Iset_{\mathrm{RM}}(K)$ be the corresponding Reed--Muller information set
      (all monomials of degree $\le r^\star$). Then there exists
      $\rho^\star<\infty$ such that for all $\rho\ge\rho^\star$,
      $\Iset_{\mathrm{rel}}(\rho,K)=\Iset_{\mathrm{RM}}(K)$ and
      \[
        \lim_{\rho\to\infty} \wmin(\rho)
        = 2^{m-r^\star},
      \]
      i.e., the reliability-based polar construction converges to the
      (truncated) RM code and its minimum distance.
\end{enumerate}
\end{theorem}

For each row-weight $d$, define
$n_d(\rho)=|\{i\in\Iset_{\mathrm{rel}}(\rho,K): D_i=d\}|$. Then, 
$\wmin(\rho)=\min\{d: n_d(\rho)>0\}$. Theorem~\ref{thm:distance-staircase}
says that, under the exponential reliability model, the functions $n_d(\rho)$
are nonincreasing in $\rho$ and vanish successively as the design SNR grows.
Each time $n_d(\rho)$ hits zero, the minimum distance jumps from $d$ to the
next larger row-weight (often $2d$), with the final plateau equal to the RM
distance. In practice, for a fixed \emph{operating} SNR, this distance gain is
accompanied by a gradual loss in aggregate SC reliability
$\sum_{i\in\Iset_{\mathrm{rel}}(\rho,K)} Z_i(\rho_{\mathrm{op}})$ as the
design SNR is pushed far above the operating point, which is precisely the
trade-off that the mixed design in this paper is intended to control.

\section{Bit-wise Mixed Metric for Near-ML Decoders}
\label{sec:sc-ml}

We now connect the mixed cost in Section~\ref{sec:mixed} to error events in
SC-based near-ML decoders, focusing on SCL decoding. 

\subsection{SC first-error events}

For SC decoding, let
\[
  \mathcal{E}_i^{\mathrm{SC}}
  =
  \{\hat u_j=u_j\ \forall j<i,\ \hat u_i\neq u_i\},\quad i\in\Iset,
\]
be the \emph{first-error} event, with probability
$P_e^{\mathrm{SC}}(i)=\Pr(\mathcal{E}_i^{\mathrm{SC}})$. The SC block error
event decomposes as
\[
  \mathcal{E}_B^{\mathrm{SC}}
  = \bigsqcup_{i\in\Iset} \mathcal{E}_i^{\mathrm{SC}},
  \quad
  P_B^{\mathrm{SC}}
  = \sum_{i\in\Iset} P_e^{\mathrm{SC}}(i).
\]

\subsection{Per-bit split of a truncated ML union bound}

Consider a truncated ML union bound
\[
  U_B^{(t)}
  = \sum_{w\in\mathcal{W}_t} A_w P_w(W),
  \quad \mathcal{W}_t\subseteq[\wmin,2\wmin].
\]
Grouping terms by bit position yields
\[
  U_B^{(t)}
  = \sum_{i\in\Iset} \Phi^{\mathrm{ML}}(i),
\]
for suitably defined $\Phi^{\mathrm{ML}}(i)$ depending on $A_w^{(i)}$ and
$\mathcal{W}_t$ (see, e.g.,~\cite{On_Partial_Weight_Distribution_of_Polar_Codes,DragoiRowshanOnePointFiveD,2504.19544v2}).

\subsection{Error-event decomposition for SCL}

For SCL decoding with list size $L$, let
$\mathcal{S}_i^{\mathrm{SCL}}(\by)$ denote the set of surviving information
prefixes at depth $i$, and $\hat\bc^{\mathrm{SCL}}(\by)$ its output. Let
$\bc_0$ be the transmitted codeword.

\begin{definition}[Pruning and ML-like events for SCL]
Define
\begin{align*}
  \mathcal{E}_{\mathrm{prune}}^{\mathrm{SCL}}
  &:= \Bigl\{\by:\ \exists i\in\Iset\ \text{s.t.}
  \bu_{0,0}^{i-1}\notin\mathcal{S}_i^{\mathrm{SCL}}(\by)\Bigr\},\\
  \mathcal{E}_{\mathrm{ML\text{-}like}}^{\mathrm{SCL}}
  &:= \Bigl\{\by:\ \bu_0\in\mathcal{S}_{|\Iset|}^{\mathrm{SCL}}(\by),\
  \hat\bc^{\mathrm{SCL}}(\by)\neq\bc_0\Bigr\},
\end{align*}
where $\bu_0$ is the true information vector.
\end{definition}

\begin{lemma}[SCL block-error decomposition]
\label{lem:SCL-e-decomp}
For SCL decoding,
\[
  \mathcal{E}_B^{\mathrm{SCL}}
  =
  \mathcal{E}_{\mathrm{prune}}^{\mathrm{SCL}}
  \,\dot\cup\,
  \mathcal{E}_{\mathrm{ML\text{-}like}}^{\mathrm{SCL}},
\]
and $\mathcal{E}_{\mathrm{ML\text{-}like}}^{\mathrm{SCL}}\subseteq \mathcal{E}_B^{\mathrm{ML}}$. Consequently,
\begin{equation}
  P_B^{\mathrm{SCL}}
  \le
  P\bigl(\mathcal{E}_{\mathrm{prune}}^{\mathrm{SCL}}\bigr)
  + P_B^{\mathrm{ML}}.
  \label{eq:SCL-prune-ML}
\end{equation}
\end{lemma}

The ML-like term in~\eqref{eq:SCL-prune-ML} can be approximated by
$U_B^{(t)}=\sum_{i\in\Iset}\Phi^{\mathrm{ML}}(i)$. The pruning term is
algorithm-dependent, but simulations and shifted-pruning analyses
(e.g.~\cite{Rowshan2019ImprovedListDecoding,Rowshan2022SCListFlipShiftedPruning})
show that pruning typically occurs after several SC-type errors along the true
path, and is therefore strongly correlated with SC first-error events.

Motivated by this, we replace $P(\mathcal{E}_{\mathrm{prune}}^{\mathrm{SCL}})$
by $\sum_{i\in\Iset} P_e^{\mathrm{SC}}(i)$ as a surrogate and obtain the
bit-wise design template
\begin{equation}
  P_B^{\mathrm{SCL}}
  \;\lesssim\;
  \sum_{i\in\Iset} P_e^{\mathrm{SC}}(i)
  + \sum_{i\in\Iset} \Phi^{\mathrm{ML}}(i)
  + R_t,
  \label{eq:SCL-bit-template}
\end{equation}
where $R_t\ge0$ accounts for truncation and overlap effects and the symbol
$\lesssim$ hides channel- and decoder-dependent constants.

\subsection{Bit-wise mixed score}

Equation~\eqref{eq:SCL-bit-template} naturally suggests ranking indices by
\begin{equation}
  \Psi(i)
  = P_e^{\mathrm{SC}}(i) + \alpha\,\Phi^{\mathrm{ML}}(i),
  \label{eq:Psi-def}
\end{equation}
for some tuning parameter $\alpha>0$. Comparing~\eqref{eq:Psi-def} with
$J_K(i)$ in~\eqref{eq:JK-def}, we see that $J_K(i)$ is a K-dependent
instantiation of this idea where $\Phi^{\mathrm{ML}}(i)$ is approximated by a
distance penalty $D_K(i)$ concentrated on maximum-degree monomials and scaled
by $Z(W)^{\wmin}$.

\begin{table*}[ht]
\centering
\caption{Representative reliability vs mixed designs on BPSK-AWGN.}
\label{tab:codes}
\footnotesize
\setlength{\tabcolsep}{4pt}
\renewcommand{\arraystretch}{0.95}
\begin{tabular}{l *{4}{cc}}
\toprule
 & \multicolumn{2}{c}{$(N,K)=(128,64)$, GA at $4$ dB}
 & \multicolumn{2}{c}{$(512,256)$, GA at $5$ dB}
 & \multicolumn{2}{c}{$(1024,512)$, GA at $3$ dB}
 & \multicolumn{2}{c}{$(32768,16384)$, GA at $5$ dB} \\
\cmidrule(lr){2-3}\cmidrule(lr){4-5}\cmidrule(lr){6-7}\cmidrule(lr){8-9}
 & $\C(\Iset_{\mathrm{rel}})$ & $\C(\Iset_{\mathrm{mix}})$
 & $\C(\Iset_{\mathrm{rel}})$ & $\C(\Iset_{\mathrm{mix}})$
 & $\C(\Iset_{\mathrm{rel}})$ & $\C(\Iset_{\mathrm{mix}})$
 & $\C(\Iset_{\mathrm{rel}})$ & $\C(\Iset_{\mathrm{mix}})$ \\
\midrule
$w_{\min}$
  & 8 & 16
  & 16 & 16
  & 16 & 32
  & 32 & 32 \\
$A_{w_{\min}}$
  & 304 & 94{,}488
  & 18{,}528 & 3{,}680
  & 2{,}752 & 12{,}673{,}632
  & 560{,}988{,}160 & 154{,}140{,}672 \\
$\sum_{i\in\Iset} P_i^{\mathrm{SC}}$
  & $2.29\!\times\!10^{-3}$ & $2.39\!\times\!10^{-2}$
  & $1.42\!\times\!10^{-7}$ & $1.59\!\times\!10^{-6}$
  & $2.69\!\times\!10^{-6}$ & $4.19\!\times\!10^{-6}$
  & $3.35\!\times\!10^{-9}$ & $3.36\!\times\!10^{-9}$ \\
$\UB_{w_{\min}}$
  & $1.3\!\times\!10^{-2}$ & $1.8\!\times\!10^{-4}$
  & $1.9\!\times\!10^{-7}$ & $3.8\!\times\!10^{-8}$
  & $1.8\!\times\!10^{-3}$ & $1.9\!\times\!10^{-7}$
  & $6.0\!\times\!10^{-14}$ & $1.6\!\times\!10^{-14}$ \\
$|\Iset_{\mathrm{rel}}\triangle\Iset_{\mathrm{mix}}|$
  & \multicolumn{2}{c}{$2\times5$}
  & \multicolumn{2}{c}{$2\times9$}
  & \multicolumn{2}{c}{$2\times7$}
  & \multicolumn{2}{c}{$2\times11$} \\
\bottomrule
\end{tabular}
\end{table*}

\section{Asymptotic Behaviour as $N\to\infty$}
\label{sec:asymptotic}

We briefly explain why the benefits of mixed constructions are mainly
finite-length phenomena. 

\subsection{Local perturbations of reliability-based sets}

Let $K=\lfloor RN\rfloor$ with $0<R<I(W)$ fixed. For each $N$, let
$\Iset_{\mathrm{rel}}(N,K)$ denote a reliability-based information set of size
$K$ (e.g., the $K$ smallest $Z_i$).

\begin{definition}[Local perturbation]
A sequence of information sets $\{\Iset(N)\}_N$ of size $K$ is a
\emph{local perturbation} of $\{\Iset_{\mathrm{rel}}(N,K)\}_N$ if
\[
  \bigl|\Iset(N)\triangle\Iset_{\mathrm{rel}}(N,K)\bigr|
  \le L(N),
\]
with $L(N)=o(N)$ as $N\to\infty$, where $\triangle$ is the symmetric difference.
\end{definition}

Intuitively, $L(N)$ is the number of positions that differ from the pure
reliability construction; mixed designs typically satisfy $L(N)\ll N$.

\subsection{Effect on SC sum and minimum-weight union bound}

By polarization, the selected bit-channels in $\Iset_{\mathrm{rel}}(N,K)$ have
$Z_i$ that decay doubly exponentially in $m=\log_2 N$. Exchanging a sublinear
number of such indices can only have a vanishing effect on the SC sum.

\begin{lemma}[Effect on SC sum]
\label{lem:local-Z}
Let $\Iset_1,\Iset_2$ be two information sets of size $K$ chosen among the
``good'' bit-channels for a BMS channel $W$, differing in at most $L(N)=o(N)$
indices. Then
\[
  \left|\sum_{i\in\Iset_1} Z_i - \sum_{i\in\Iset_2} Z_i\right|
  \xrightarrow[N\to\infty]{} 0.
\]
\end{lemma}

For decreasing monomial codes with fixed maximum degree $r$, recall that
$\wmin=2^{m-r}$ and $A_{\wmin}(\Iset)$ is given by~\eqref{eq:Awmin-ir-sum}. A
local change in the set of degree-$r$ monomials perturbs $A_{\wmin}$ by at
most $O(L(N))$, and the associated minimum-weight contribution to the union
bound by at most $O(L(N)\gamma(W)^{2^{m-r}})$.

\begin{lemma}[Effect on truncated minimum-weight term]
\label{lem:UB-local}
Let $\UB_{\wmin}(\Iset)=A_{\wmin}(\Iset)\,\gamma(W)^{\wmin}$ be the
minimum-weight part of the ML union bound. If $\Iset_1,\Iset_2$ are decreasing
monomial sets of size $K$ with the same maximum degree $r$, differing in at
most $L(N)=o(N)$ monomials, then
\[
  \bigl|
    \UB_{\wmin}(\Iset_1) - \UB_{\wmin}(\Iset_2)
  \bigr|
  \xrightarrow[N\to\infty]{} 0.
\]
\end{lemma}


Combining Lemmas~\ref{lem:local-Z} and~\ref{lem:UB-local} shows that any $k$-dependent mixed perturbation of a reliability-based information set has asymptotically negligible effect on SC and ML union bounds. Thus, mixed design mainly benefits finite lengths: as $N$ grows, polarization dominates and the added weight term affects only constant factors.

\section{Numerical Results}
\label{sec:numerical}

\begin{figure}
    \centering
    \includegraphics[width=1\linewidth]{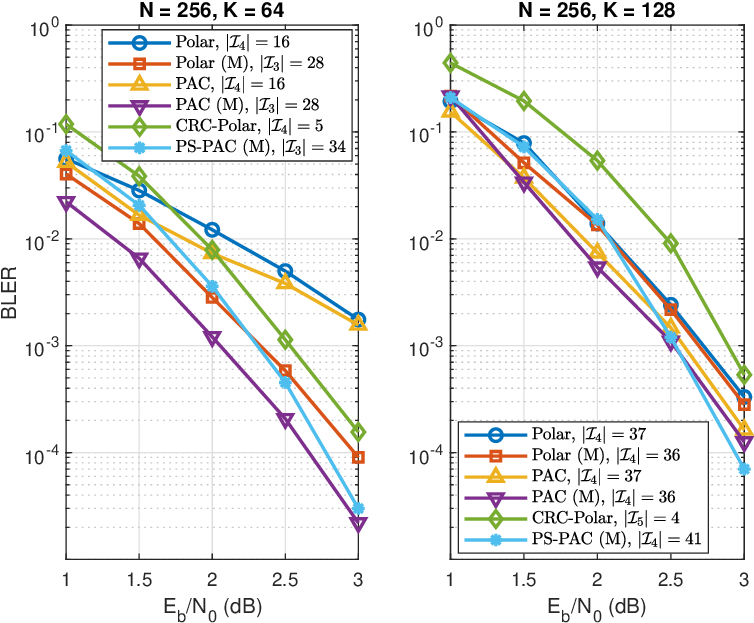}
    \caption{BLER of codes obtained by mixed metric (M). }
    \label{fig:BLER_mixed}
\end{figure}

We illustrate the trade-off between pure reliability-based and mixed designs
for several $(N,K)$ pairs over BPSK-AWGN. All designs use GA estimates of
$P_i^{\mathrm{SC}}$ and closed-form weight contributions at fixed maximum
degree $r$. 
Table~\ref{tab:codes} reports structural and union-bound metrics for several
lengths, comparing the pure reliability-based polar code
$\C(\Iset_{\mathrm{rel}})$ and the mixed design $\C(\Iset_{\mathrm{mix}})$.
The quantity $\UB_{\wmin}$ is the truncated minimum-weight ML union bound
$A_{\wmin} Z(W)^{\wmin}$ at the design SNR, and
$|\Iset_{\mathrm{rel}}\triangle\Iset_{\mathrm{mix}}|$ measures how many
positions differ between the two information sets. 
This table highlights the following points: 
  (a) at short and moderate lengths, the mixed design can double $\wmin$ or dramatically reduce $A_{\wmin}$ while incurring only a moderate increase in $\sum P_i^{\mathrm{SC}}$; the minimum-weight term $\UB_{\wmin}$ improves by one to several orders of magnitude;
  (b) at large lengths, the reliability sums become extremely small and almost identical across constructions, while multiplicity changes mainly affect constant factors in the ML union bound, consistent with the asymptotic analysis of Section~\ref{sec:asymptotic}.

Fig.~\ref{fig:BLER_mixed} compares polar, PAC, CRC-polar, and PS-PAC codes of length $N=2^m=256$ and rates $R=1/4$ and $1/2$, designed by GA at 3 and 4 dB, with their mixed-metric counterparts (M). We use $\mathbf{p}=[1\,0\,1\,1\,0\,1\,1]$ for PAC pretransformation \cite{arikan2}, a 12-bit CRC with polynomial $\mathrm{0xC06}$ for CRC-polar coding, and $\alpha=6$ for PS-PAC \cite{gu2025pac}. As indicated by $|\mathcal{I}_r|$ in the legends, the (256,64) mixed-metric designs exclude all degree-$4$ monomials (i.e., $\mathbf{G}_N$-rows of weight $2^{m-4}=16$), increasing the code distance to $\ge 32$ \cite{rowshan2023minimum} and yielding performance governed by $A_{32}$ at high $E_b/N_0$. Because the PS-PAC code has a larger $|\mathcal{I}_3|$, it employs more low-reliability sub-channels and hence under-performs at low SNR regime, but surpasses polar (M) and PAC (M) at high SNR. 
A similar trend appears for (256,128), except that the CRC-polar code includes $\mathbf{G}_N$-rows of weight $2^{m-5}=8$, which affects its overall reliability and ($w_{\min},A_{w_{\min}}$). 
Due to small difference in $|\mathcal{I}_4|$ of codes, the gain is insignificant, except for the PS-PAC code due to significant reduction in $A_{w_{\min}}$. Note that reliability-based PS-PAC code have $|\mathcal{I}_5|=3$, yielding poor BLER.

\section{Summary and Future Directions}
\label{sec:conclusion}
We introduced a mixed reliability–weight metric for constructing polar codes  (software is available in \cite{rowshan2026mixedGITHUB}). The method penalizes only maximum‑degree monomials using closed‑form orbit counts of minimum‑weight codewords, scaled by a Bhattacharyya factor and combined with SC‑based reliability into a per‑bit cost. This mixed metric connects naturally to bit‑wise SC decomposition and truncated ML union bounds for SCL decoding, yielding designs that act as small perturbations of standard reliability constructions. Future work includes extending the metric to PW‑based sequences and embedding the weight term into universal partial orders for channel‑independent, distance‑aware polar designs.

\clearpage
\bibliographystyle{IEEEtran}
\bibliography{refs}

\clearpage
\appendices

\section{Proof of Lemma~\ref{lem:negligible-high-weight}}
\label{app:proof-ml-negl}

Let $w_0=w_{\min}(i)$ and $\gamma=\gamma(W)$. By the exponential bound on
$P_w(W)$ and the definition of $\Phi^{\mathrm{ML}}(i)$,
\[
  \Phi^{\mathrm{ML}}(i)
  \le
  c(W)\sum_{w\ge w_0} A_w^{(i)} \gamma^{w}
  =
  c(W)\,\gamma^{w_0}
  \sum_{w\ge w_0} A_w^{(i)} \gamma^{w-w_0}.
\]
For fixed rate $R$, there are at most $2^{RN}$ codewords, so
$\sum_{w\ge w_0}A_w^{(i)}\gamma^{w-w_0}\le C_{\mathrm{up}}(W,R)$ for some
constant. Thus
\[
  \Phi^{\mathrm{ML}}(i)
  \le
  C_{\mathrm{up}}(W,R)\,\gamma^{w_{\min}(i)}.
\]
If $\deg(f_i)\le r-1$, then $w_{\min}(i)\ge 2\wmin$, so
$\Phi^{\mathrm{ML}}(i)\le C_{\mathrm{up}}(W,R)\,\gamma^{2\wmin}$.

For the degree-$r$ index $j$ with $w_{\min}(j)=\wmin$ and
$A_{\wmin}^{(j)}\ge1$,
\[
  \Phi^{\mathrm{ML}}(j)
  \ge
  A_{\wmin}^{(j)} P_{\wmin}(W)
  \ge
  c_{\mathrm{low}}(W)\,\gamma^{\wmin}
\]
for some $c_{\mathrm{low}}(W)>0$. Hence
\[
  \frac{\Phi^{\mathrm{ML}}(i)}{\Phi^{\mathrm{ML}}(j)}
  \le
  \frac{C_{\mathrm{up}}(W,R)}{c_{\mathrm{low}}(W)}\,\gamma^{\wmin}
  \xrightarrow[m\to\infty]{} 0,
\]
because $\gamma\in(0,1)$ and $\wmin=2^{m-r}\to\infty$.

\section{Proof of Lemma~\ref{lem:SCL-e-decomp}}
\label{app:proof-scl}

Fix the transmitted codeword $\bc_0$ and its information vector $\bu_0$. For
each $\by$, SCL produces a list
$\mathcal{S}_{|\Iset|}^{\mathrm{SCL}}(\by)$ and outputs
$\hat\bc^{\mathrm{SCL}}(\by)$ with maximal path metric among candidates in the
list.

\emph{Decomposition.} If $\hat\bc^{\mathrm{SCL}}(\by)=\bc_0$, then no SCL
error occurs and $\by\notin\mathcal{E}_B^{\mathrm{SCL}}$. In particular, the
true path is never pruned and $\bu_0\in\mathcal{S}_{|\Iset|}^{\mathrm{SCL}}(\by)$, so
$\by\notin\mathcal{E}_{\mathrm{prune}}^{\mathrm{SCL}}$ and
$\by\notin\mathcal{E}_{\mathrm{ML\text{-}like}}^{\mathrm{SCL}}$.

If $\hat\bc^{\mathrm{SCL}}(\by)\neq\bc_0$, then either:
(i) the true prefix $\bu_{0,0}^{i-1}$ leaves the list at some depth $i$, i.e.
$\by\in\mathcal{E}_{\mathrm{prune}}^{\mathrm{SCL}}$; or
(ii) all true prefixes survive to depth $|\Iset|$, so
$\bu_0\in\mathcal{S}_{|\Iset|}^{\mathrm{SCL}}(\by)$, but SCL still chooses a
different codeword, i.e.\ $\by\in\mathcal{E}_{\mathrm{ML\text{-}like}}^{\mathrm{SCL}}$.
These two events are mutually exclusive and cover all SCL errors, yielding
\[
  \mathcal{E}_B^{\mathrm{SCL}}
  =
  \mathcal{E}_{\mathrm{prune}}^{\mathrm{SCL}}
  \,\dot\cup\,
  \mathcal{E}_{\mathrm{ML\text{-}like}}^{\mathrm{SCL}}.
\]

\emph{Inclusion in ML error.} If
$\by\in\mathcal{E}_{\mathrm{ML\text{-}like}}^{\mathrm{SCL}}$, then
$\bu_0\in\mathcal{S}_{|\Iset|}^{\mathrm{SCL}}(\by)$ and
$\hat\bc^{\mathrm{SCL}}(\by)\neq\bc_0$. The SCL metric is (a monotone function
of) the log-likelihood $\Lambda(\bc,\by)$, so SCL selects some
$\tilde\bc\neq\bc_0$ with
$\Lambda(\tilde\bc,\by)>\Lambda(\bc_0,\by)$ (or ties are broken away from
$\bc_0$). The ML decoder maximises the same metric over all codewords in
$\C(\Iset)$ and cannot choose $\bc_0$ when a strictly better $\tilde\bc$
exists. Hence the ML decoder also errs and
$\by\in\mathcal{E}_B^{\mathrm{ML}}$, proving
$\mathcal{E}_{\mathrm{ML\text{-}like}}^{\mathrm{SCL}}\subseteq\mathcal{E}_B^{\mathrm{ML}}$.
Taking probabilities gives~\eqref{eq:SCL-prune-ML}.


\section{Proof of Lemma~\ref{lem:local-Z}}

Fix $\beta\in(0,1/2)$. By polarization, for any $\varepsilon>0$ and all
sufficiently large $m$ there exists a set of ``good'' indices
$\mathcal{G}_\varepsilon(N)$ with $|\mathcal{G}_\varepsilon(N)|\approx I(W)N$
such that $Z_i\le 2^{-2^{\beta m}}$ for all $i\in\mathcal{G}_\varepsilon(N)$.
We assume both $\Iset_1$ and $\Iset_2$ are subsets of $\mathcal{G}_\varepsilon(N)$.

Let $\Delta_Z(N)$ be as in the lemma. Then
\[
  \Delta_Z(N)
  \le
  \sum_{i\in\Iset_1\triangle\Iset_2} Z_i
  \le
  L(N)\cdot 2^{-2^{\beta m}}.
\]
Since $L(N)=o(N)$ and $N=2^m$, there exists $\alpha<1$ such that
$L(N)\le 2^{\alpha m}$ for large $m$, giving
\[
  \Delta_Z(N)
  \le
  2^{\alpha m-2^{\beta m}}
  \xrightarrow[m\to\infty]{} 0.
\]
This proves the claim.

\section{Proof of Lemma~\ref{lem:UB-local}}

Recall
\[
  A_{\wmin}(\Iset)
  =
  \sum_{f\in\Ir(\Iset)} 2^{r+|\lambda_f|}
  =
  \sum_{i\in\Iset} C_K(i).
\]
Since $r$ is fixed and $|\lambda_f|\le r(m-r)$, there exists a constant
$C_r>0$ such that $2^{r+|\lambda_f|}\le C_r$ for all degree-$r$ monomials and
all $m$. If $\Iset_1$ and $\Iset_2$ differ in at most $L(N)$ monomials, then
they differ in at most $L(N)$ degree-$r$ monomials, and each such change
perturbs $A_{\wmin}$ by at most $C_r$. Thus
\[
  \bigl|A_{\wmin}(\Iset_1)-A_{\wmin}(\Iset_2)\bigr|
  \le C_r L(N).
\]
Multiplying by $\gamma(W)^{\wmin}$ with $\wmin=2^{m-r}$ yields
\[
  \bigl|
    \UB_{\wmin}(\Iset_1)-\UB_{\wmin}(\Iset_2)
  \bigr|
  \le
  C_r L(N)\,\gamma(W)^{2^{m-r}},
\]
which tends to zero since $\gamma(W)\in(0,1)$ and $L(N)=o(N)$.

\section{Proof of Theorem~\ref{thm:distance-staircase}}

\begin{proof}
We only need the exponential model~\eqref{eq:exp-model}. For (i), fix
$0\le\rho_1<\rho_2$ and suppose by contradiction that
$\wmin(\rho_2) < \wmin(\rho_1)$. Let
$S_1=\Iset_{\mathrm{rel}}(\rho_1,K)$,
$S_2=\Iset_{\mathrm{rel}}(\rho_2,K)$, and set
$d_1=\min_{i\in S_1} D_i$, $d_2=\min_{i\in S_2} D_i$. The assumption gives
$d_2<d_1$. Then there exists $j\in S_2$ with $D_j=d_2$, while for every
$i\in S_1$ we have $D_i\ge d_1>d_2$, hence $D_i>D_j$.

For each such pair $(i,j)$, by~\eqref{eq:exp-model},
\[
  \frac{Z_i(\rho)}{Z_j(\rho)}
  = \frac{a_i}{a_j} e^{-c(D_i-D_j)\rho},
\]
which is strictly decreasing in $\rho$ because $D_i>D_j$. At $\rho_1$,
$j\notin S_1$ while $S_1$ contains the $K$ smallest $Z_i(\rho_1)$, so there
exists some $i\in S_1$ with $Z_i(\rho_1)\le Z_j(\rho_1)$, i.e.,
$Z_i(\rho_1)/Z_j(\rho_1)\le1$. Since the ratio is decreasing in $\rho$, we
get $Z_i(\rho)<Z_j(\rho)$ for all $\rho>\rho_1$, in particular at $\rho_2$.
Thus, at $\rho_2$ each such $i\in S_1$ is \emph{more} reliable than $j$, so
if $j\in S_2$ at least one of those $i$ must also belong to $S_2$ (otherwise
$S_2$ would not contain the $K$ smallest $Z_i(\rho_2)$), and in particular
the minimal row-weight in $S_2$ could not be less than $d_1$. This contradicts
$d_2<d_1$, so we must have $\wmin(\rho_2)\ge\wmin(\rho_1)$.

The piecewise-constant staircase structure follows because for each pair
$i\ne j$ with $D_i\ne D_j$ the equation $Z_i(\rho)=Z_j(\rho)$ has at most one
solution in $\rho$, so the global ordering of $\{Z_i(\rho)\}_i$ can change
only at finitely many $\rho$ in any bounded interval; between such points
$\Iset_{\mathrm{rel}}(\rho,K)$ and hence $\wmin(\rho)$ are constant. Each jump
occurs exactly when the current minimal row-weight $d$ disappears from
$\Iset_{\mathrm{rel}}(\rho,K)$ and the minimum switches to the next larger
row-weight among the selected indices.

For (ii), if $D_i>D_j$ then
\[
  \frac{Z_i(\rho)}{Z_j(\rho)}
  = \frac{a_i}{a_j} e^{-c(D_i-D_j)\rho}
  \xrightarrow[\rho\to\infty]{} 0,
\]
so for sufficiently large $\rho$ we have $Z_i(\rho)<Z_j(\rho)$ whenever
$D_i>D_j$. Thus, beyond some $\rho^\star$ the ordering of all $Z_i(\rho)$ by
increasing value coincides with the ordering by decreasing row-weight $D_i$,
and $\Iset_{\mathrm{rel}}(\rho,K)$ must consist of the $K$ indices with
largest $D_i$, i.e., the Reed--Muller information set $\Iset_{\mathrm{RM}}(K)$
(up to ties). In the decreasing-monomial representation, this corresponds to
the RM code $R(r^\star,m)$ (truncated if $K<\dim R(r^\star,m)$), whose
minimum distance is $2^{m-r^\star}$. Since $\wmin(\rho)$ is nondecreasing and
bounded above by $2^{m-r^\star}$, we obtain
$\lim_{\rho\to\infty}\wmin(\rho)=2^{m-r^\star}$.
\end{proof}


\end{document}